\documentclass[aps,reprint,floatfix,superscriptaddress]{revtex4-1}
\usepackage[english]{babel}
\usepackage[utf8]{inputenc}
\usepackage{bbold}
\usepackage{epsfig}
\usepackage{amsmath}
\usepackage{units}
\usepackage[colorlinks]{hyperref}
\usepackage{enumerate}
\usepackage{color}		  				
\usepackage{lipsum}

\begin{document}

\title{Magnetic impurities along the edge of a quantum spin Hall insulator:\\ Realizing a one-dimensional AIII insulator}
\author{G.A.R. van Dalum}
\affiliation{Institute for Theoretical Physics and Center for Extreme Matter and Emergent Phenomena,
Utrecht University, Princetonplein 5, 3584 CC Utrecht, The Netherlands}

\author{C. Ortix}
\affiliation{Institute for Theoretical Physics and Center for Extreme Matter and Emergent Phenomena,
Utrecht University, Princetonplein 5, 3584 CC Utrecht, The Netherlands}
\affiliation{Dipartimento di Fisica ``E. R. Caianiello'',
Universit\`a di Salerno, I-84084 Fisciano (Salerno), Italy }

\author{L. Fritz}
\affiliation{Institute for Theoretical Physics and Center for Extreme Matter and Emergent Phenomena,
Utrecht University, Princetonplein 5, 3584 CC Utrecht, The Netherlands}

\begin{abstract}
In this paper we construct a one-dimensional insulator with an approximate chiral symmetry belonging to the AIII class and discuss its properties. The construction principle is the intentional pollution of the edge of a two-dimensional quantum spin Hall insulator with magnetic impurities. The resulting bound states hybridize and disperse along the edge. We discuss under which circumstances this chain possesses zero-dimensional boundary modes on the level of an effective low-energy theory. The main appeal of our construction is the independence on details of the impurity lattice: the zero modes are stable against disorder and random lattice configurations. We also show that in the presence of Rashba coupling, which changes the symmetry class to A, one can still expect localized half-integer boundary excess charges protected by mirror symmetry although there is no nontrivial topological index.  All of the results are confirmed numerically in a microscopic model.
\end{abstract}

\maketitle

\section{Introduction}

The vast majority of topological phases of matter are characterized by the presence of metallic boundary modes directly linked to a topological invariant~\cite{has10,qi11}. The
Chern number, for instance, counts the number of chiral states appearing at the edge of quantum Hall insulators~\cite{tho82,hal88,kli80}. Likewise, the presence of  an odd number of Kramers' related pairs of edge bands in quantum spin Hall insulators~\cite{kan05,kan05b,ber06,mol07} is in a one-to-one correspondence with the nontrivial value of a ${\mathbb Z}_2$  invariant.

In crystalline systems with an additional set of spatial symmetry, additional topological crystalline phases appear~\cite{fu11,hsi12,slager2013,liu14b,hsi14,kruthoff2017}.  In these systems, a nontrivial value of the bulk topological invariant is manifested in gapless boundary modes that violate a stronger version of the fermion doubling theorem~\cite{fan17,kha18}. Most importantly, the metallic boundary modes appear only on surfaces that are left invariant under the protecting symmetry. Crystalline symmetries also lead to the presence of different topologically nontrivial phases, dubbed as higher-order topological insulators~\cite{sch18,ben17,ben17-2,sch18b,gei18,pet18,ser18,kha18b,lan17,sit12,son17,imh18,mie18,koo18}, characterized by conventional gapped surface states, but with metallic chiral or helical modes at the hinges connecting surfaces related by the protecting crystalline symmetry.

Topological crystalline phases cannot appear in one-dimensional systems for the very simple reason that any spatial symmetry interchanges the ends of the chain.
Indeed, the zero-energy metallic modes of a one-dimensional topological insulating phase necessitate the presence of an internal chiral or particle-hole symmetry.
Nevertheless, the ground state of a one-dimensional insulator can be characterized by gauge-invariant topological indices even
in the complete absence of internal symmetries.
In one-dimensional systems, in particular,
these topological indices govern the excess of electric charge localized at the ends of the atomic chain~\cite{mie17}, which are (fractionally) quantized when additional crystalline symmetries are present~\cite{mie17,lau16}. Quantized excess charges have recently been proposed to appear in newly synthesized linear chains of a transitional metal dichalcogenide~\cite{pha20}.

In this work we engineer a one-dimensional system with nontrivial topological index by means of intentional ``pollution'' of a host system. The usual starting point is a nontrivial host system on whose boundary adatoms are deposited in a regular manner. These adatoms or ``impurities'' bind electronic in-gap states which can hybridize. This leads to impurity bands with potentially topological character~\cite{choy,np,brau1,klin,vazifeh,pientka2,poyh,heimes,reis,bry,west,balents,kimme,minarelli}. A recent example of this more generic approach is the deposition of ferromagnetic atoms in a chain on superconductors. These magnetic impurities individually bind a pair of Shiba bound states. Once these bound states hybridize, one obtains artificial wires which have been shown to potentially host topological bands. This was, for instance, demonstrated in an experiment where iron atoms were arranged in a chain structure on a superconductor. These efforts have led to the observation of Majorana-like signatures at the ends of the chain~\cite{np2,feldman,ruby,pawlak}. In a related experiment~\cite{ruby2} using Co atoms these modes were absent, meaning that the protocol does not seem robust and independent of details. Theoretically, this scheme has also been extended to higher dimensions~\cite{nakosai}, for instance leading to two-dimensional topological superconductors with a wide range of notably high Chern numbers~\cite{ront1,ront2,kaladzhyan}.

For concreteness, we here consider the boundary of a two-dimensional quantum spin Hall insulator where the boundary modes are gapped due to the application of a magnetic field. Subsequently, we deposit magnetic impurities onto the boundary. These magnetic impurities bind electrons in bound states. If the distance between the impurities is not too large, these impurity bound states overlap and hybridize, effectively leading to a one-dimensional hopping problem. We demonstrate that such a setup can realize a one-dimensional system in the AIII symmetry class of the Altland-Zirnbauer~\cite{AltlandZirnbauer} classification. It possesses an integer topological index and half-integer boundary excess charges, stable against disorder. This is shown explicitly, both analytically and numerically.
The paper is organized as follows: we start by introducing the setup in Sec.~\ref{sec:model}. In Sec.~\ref{sec:dilute} we discuss an approximate analytical solution to the impurity problem. This includes an argument for the stability of these boundary modes against disorder. In Sec.~\ref{sec:numerical} we discuss the setup numerically on the level of a microscopic lattice model, and we end with the conclusions in Sec.~\ref{sec:conclusion}.

\section{Microscopic model}\label{sec:model}
\begin{figure}[t]
\centering
\includegraphics[width=0.45\textwidth]{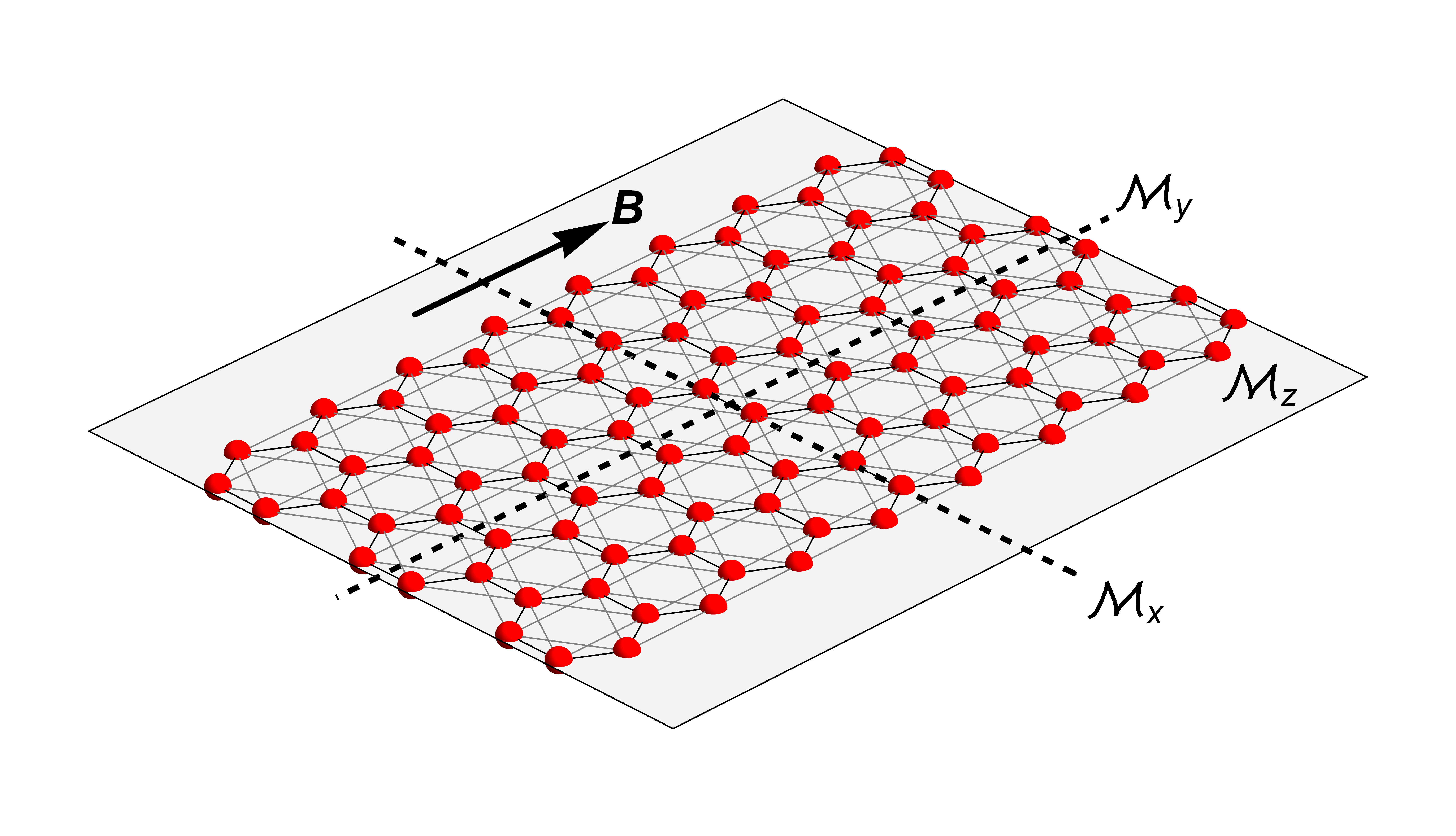}
\caption{\label{fig:setupsymm}Illustration of the host system $H_\text{KM}+H_\text{mag}$ and the mirror plane and lines. In the case of periodic boundary conditions in the $\hat{\mathbf{x}}$ direction, there is an additional mirror line $\mathcal{M}_x$ for each discrete translation of $\sqrt{3}/2$ times the nearest neighbor distance. Note that $\mathcal{M}_z$ and $\mathcal{M}_y$ are broken by the magnetic field $\mathbf{B}$, while $\mathcal{M}_x$ and $\mathcal{M}_{z}\Theta$ are preserved.}
\end{figure}

In order to study the topological properties of impurity bound states on the gapped edge of a quantum spin Hall insulator, we introduce the Hamiltonian
\begin{equation}
H=H_\text{KM}+H_\text{mag}+H_\text{imp}.\label{eq:Hfull}
\end{equation}
The itinerant electrons live on a honeycomb lattice (point group ${\mathcal C}_{6v}$) and are described by the Kane-Mele model~\cite{kan05b},
\begin{equation}
H_\text{KM}=-t\sum_\alpha\sum_{\langle i,j\rangle}c_{i\alpha}^\dagger c^{\phantom{\dagger}}_{j\alpha}+it_2\sum_{\alpha\beta}\sum_{\langle\langle i,j\rangle\rangle}\nu_{ij}c_{i\alpha}^\dagger\sigma_{\alpha\beta}^zc^{\phantom{\dagger}}_{j\beta},\label{eq:HKM}
\end{equation}
where $c_{i\alpha}^\dagger$ ($c_{i\alpha}$) creates (annihilates) an electron with spin $\alpha\in\{\uparrow,\downarrow\}$ at lattice site $i\in\{1,\ldots,N\}$. The two sums involving lattice sites go over nearest neighbor pairs, $\langle i,j\rangle$, and next-nearest neighbors, $\langle\langle i,j\rangle\rangle$, respectively, with real hopping parameters $t$ and $t_2$. Moreover, $\nu_{ij}=+1$ $(-1)$ if the next-nearest neighbor path $i\rightarrow j$ is counterclockwise (clockwise), while $\sigma^z$ is the third Pauli matrix. This intrinsic spin-orbit coupling reduces the $\mathcal{SU}(2)$ spin symmetry to a residual ${\mathcal U}(1)$ spin symmetry ensured by the mirror plane symmetry ${\mathcal M}_z$. As was shown by Kane and Mele, the above model realizes a quantum spin Hall insulator at half-filling and consequently supports gapless helical edge states protected by time-reversal symmetry. Without loss of generality, in the following we will consider the helical edge states realized in the $\hat{\mathbf{x}}$ zigzag direction of the honeycomb lattice. Next, we gap these edge states by applying a uniform planar magnetic field $\mathbf{B}$ in the $\hat{\mathbf{x}}$ direction, leading to a Zeeman coupling
\begin{equation}
H_\text{mag}=B\sum_{\alpha\beta}\sum_ic_{i\alpha}^\dagger\sigma_{\alpha\beta}^xc_{i\beta}.
\end{equation}
Note that this term explicitly breaks the time-reversal symmetry, thus changing the Altland-Zirnbauer~\cite{AltlandZirnbauer} symmetry class of the model from AII to A. Furthermore, besides the threefold rotation symmetry ${\mathcal C}_3$, the planar magnetic field breaks both the mirror plane symmetry ${\mathcal M}_z$ as well as the mirror line ${\mathcal M}_y$ while preserving the orthogonal mirror line ${\mathcal M}_x$. As will become clear below, it is important to note that the combined ${\mathcal M}_{z} \Theta$ symmetry -- $\Theta$ representing the time-reversal symmetry operator -- is still preserved. The setup corresponding to the above model, the mirror plane $\mathcal{M}_z$ and the mirror lines $\mathcal{M}_{x,y}$ are shown in Fig.~\ref{fig:setupsymm}. Finally, a set of pointlike magnetic impurities, parametrized by $\tilde{V}_M$, is included on certain edge lattice sites $\{l\}\subset\{i\}$, such that
\begin{equation}
H_\text{imp}=\tilde{V}_M\sum_{\alpha\beta}\sum_lc_{l\alpha}^\dagger\sigma_{\alpha\beta}^xc^{\phantom{\dagger}}_{l\beta}.\label{eq:Himplatt}
\end{equation}

The properties of the impurity bound states can now in principle be studied numerically. Working with the full lattice model is most feasible when the lattice size remains small, which requires a dense packing of impurities. In the following, we will first study the dilute continuum limit analytically, and subsequently study the lattice limit numerically.

\section{Analytical investigation of the dilute limit}\label{sec:dilute}
\begin{figure}[t]
\centering
\includegraphics[width=0.45\textwidth]{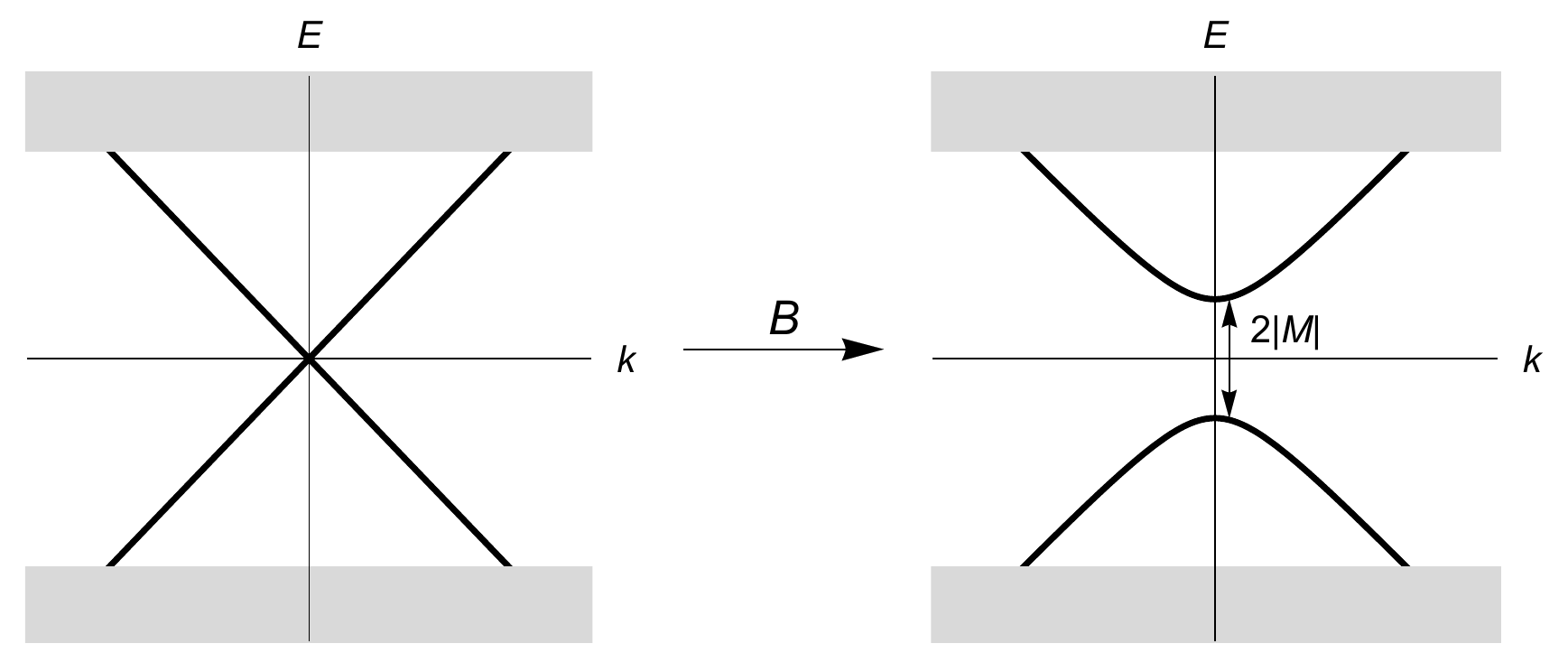}
\caption{\label{fig:dispersion_sketch}Sketch of the dispersion relation corresponding to Eq.~(\ref{eq:Hfull}) with zigzag edges and in absence of impurities. The gray regions represent the bulk bands, while the in-gap energies are associated with edge states. The left panel corresponds to $H_\text{KM}$, while the right panel illustrates the effects of introducing a planar magnetic field.}
\end{figure}

We start by deriving an effective low-energy model for the gapped helical edge modes of our quantum spin Hall insulator subject to a planar magnetic field, and will consider a single edge in isolation realizing an effective anomalous one-dimensional system. This effective model must respect the symmetries of the microscopic model, while reproducing the edge dispersion, see Fig.~\ref{fig:dispersion_sketch}. First, we notice that as long as the planar magnetic field is parallel to the edge axis, say $\hat{\mathbf{x}}$, the single edge is invariant under the mirror line symmetry ${\mathcal M}_x$. This is different for a planar magnetic field perpendicular to the edge axis as the only surviving mirror symmetry (${\mathcal M}_y$) interchanges the two edges. We can use the ${\mathcal M}_{x}$ and ${\mathcal M}_z \Theta$ symmetry constraints to write the one-dimensional effective Hamiltonian. Specifically, the mirror line symmetry reads ${\mathcal M}_x= i \sigma^x$ in spin space whereas the combined antiunitary symmetry is given by ${\mathcal M}_z \Theta= i \sigma^x {\mathcal K}$, where we have used that ${\mathcal M}_z = i \sigma^z$ and $\Theta= i \sigma^y {\mathcal K}$. Using that the Hamiltonian in spin space can be written as $\hat{\mathcal H}=\sum_{j=0,x,y,z} h_j(k) \sigma^j$, $\sigma^0$ being the identity, we have that the mirror line symmetry forces $h_{y,z}(k)$ to be odd in momentum $k$ while $h_{0,x}(k)$ has to be even. On the other hand,  the antiunitary ${\mathcal M}_z \Theta$ symmetry forces all functions to be even in momentum except for $h_z(k)$. These symmetry constraints consequently guarantee $h_y(k) \equiv 0$. Furthermore, we can also impose the function $h_0(k)$ to be zero using the following argument. Besides an irrelevant rigid shift of the energies, a sizable quadratic term $\propto k^2$ would generically violate the prime physical property of the edge states of a quantum spin Hall insulator: for any given Fermi energy an odd number of Kramers' pairs have to be intersected. As a result, we have that the system inherits an approximate chiral symmetry which is stipulated by the anticommutation relation $\left\{ \mathcal{H} , \sigma^y \right\} =0$ for all momentum values, and consequently implies that in the presence of a planar magnetic field  the system belongs to the AIII class of the Altland-Zirnbauer classification. For small momenta, an appropriate minimal model for a single edge of the clean system (\textit{i.e.}, in absence of $H_\text{imp}$) is thus given by the Hamiltonian
\begin{equation}
\hat{\mathcal{H}}_\text{min}(k)=\hbar v_Fk\sigma^z+M\sigma^x,\label{eq:Hmin}
\end{equation}
where $v_F$ is the Fermi velocity, and $M$ is the effective mass induced by the (small) external magnetic field $B$. The energy spectrum of this Hamiltonian is given by $\epsilon(k)=\pm\sqrt{\hbar^2v_F^2k^2+M^2}$, such that there is an edge gap with gap size $\Delta_\text{edge}=2|M|$. Note that the $2\times 2$ matrix structure corresponds to the actual spin of the electrons, and that the hat notation has been introduced to indicate this matrix structure. In this continuum language, the real-space impurity Hamiltonian furthermore takes the form
\begin{equation}
\hat{\mathcal{H}}_\text{imp}(x)=V_M\sum_l\delta(x-x_l)\sigma^x.\label{eq:Himp}
\end{equation}
Here, $x_l$ are the one-dimensional (1D) coordinates of the impurities along the edge, and $V_M=a_0\tilde{V}_M$, with $a_0$ being the lattice constant of the underlying honeycomb lattice (\textit{i.e.}, $\sqrt{3}$ times the nearest neighbor distance). We thus specifically place the impurities on sites on which the edge states are localized. By doing so, we effectively ignore any sublattice structure of the impurity potential. However, since the edge states are strongly localized, the neglected component of the impurity potential only modifies the bulk dispersion, such that ignoring the sublattice structure will not qualitatively change our results. The impurity bound states of this effective boundary model can be found by solving the Schr\"odinger equation in integral form, given by
\begin{equation}
\psi(x)=\int\limits_{-\infty}^\infty\mathrm{d}x^\prime\,\hat{\mathcal{G}}_0^+(x,x^\prime;E)\hat{\mathcal{H}}_\text{imp}(x^\prime)\psi(x^\prime),\label{eq:Schr}
\end{equation}
for a given bound state energy $E$, where the translationally invariant retarded Green function $\hat{\mathcal{G}}_0^+(x,x^\prime;E)=\hat{\mathcal{G}}_0^+(x-x^\prime;E)$ is the inverse Fourier transform of the $2\times 2$ matrix
\begin{equation}
\hat{\mathcal{G}}_0^+(k;E)=\big[(E+i0^+)\mathbb{1}-\hat{\mathcal{H}}_\text{min}(k)\big]^{-1}.\label{eq:G0k}
\end{equation}

Contrary to the microscopic model, the above boundary model is subject to a couple of physical limitations. In particular, this minimal low-energy model is only valid for energy scales that are small with respect to the bulk gap, while it assumes that the clean system acts as an inert continuous background to the impurities. As such, it describes the situation in which the typical distance between neighboring impurities $a_\text{imp}$ is much larger than the lattice constant of the original microscopic model $a_0$, and it is necessary to verify that all energies remain small at all times. The effective model is therefore indeed particularly appropriate for studying the limit in which the impurities are dilute. For the remainder of this section, we thus assume that the edge gap $\Delta_\text{edge}=2|M|$ is small compared to the bulk gap $\Delta_\text{bulk}\sim |t|$, and that the clean system can indeed be approximated as a continuous background to the impurities. Our strategy for studying this model is to develop an effective tight-binding model consisting of the impurity bound states described by Eq.~(\ref{eq:Schr}), and investigate the topological properties of the resulting bound state bands. In order to construct the desired tight-binding model, we first look at the bound states associated with a single isolated impurity, then include the hybridization of several of such bound states.

\subsection{Effective low-energy tight-binding model}
The central object for studying the impurity bound states is the retarded Green function in position space. Performing an inverse Fourier transformation on Eq.~(\ref{eq:G0k}), we find that this object is given by
\begin{align}
\hat{\mathcal{G}}_0^+(&x,x^\prime;E)=-\frac{1}{2\hbar v_F}\nonumber\\
&\times\left[\frac{\xi}{\hbar v_F}(E\mathbb{1}+M\sigma^x)-i\,\text{sgn}(\Delta x)\sigma^z\right]e^{-|\Delta x|/\xi}
\end{align}
for in-gap energies $|E|<|M|$. Here, $\Delta x=x-x^\prime$ is the relative 1D coordinate, the sign function has been defined with $\text{sgn}(0)\equiv 0$, and $\xi=\hbar v_F/\sqrt{M^2-E^2}$ is the correlation length of the clean boundary model. For future reference, we also define $\xi_0$ as the zero-energy correlation length, \textit{i.e.}, $\xi_0\equiv\hbar v_F/|M|=2\hbar v_F/\Delta_\text{edge}$.

In the case of a single impurity, the Schr\"odinger equation~(\ref{eq:Schr}) dictates that the bound state energy $E$ and the corresponding wave function at the impurity site $\psi(x_\text{imp})$ must satisfy the equation
\begin{equation}
\big[\mathbb{1}-V_M\hat{\mathcal{G}}_0^+(0;E)\sigma^x\big]\psi(x_\text{imp})=0.
\end{equation}
Taking the determinant of the left-hand side matrix and equating it to zero, we find two bound state energies within the domain $|E|<|M|$:
\begin{equation}
E_\pm=\pm M\left(1-\frac{2V_M^2}{V_M^2+4\hbar^2v_F^2}\right),
\end{equation}
provided that $V_M$ is finite. These energies can be tuned to be small (\textit{i.e.}, close to the center of the edge gap) by setting $|V_M|\sim 2\hbar v_F$.

Next, we consider a set of impurities with typical nearest neighbor distance $a_\text{imp}$. In the dilute limit $a_\text{imp}\gg\xi$, the bound state associated with each impurity can be approximated by the above solution of an isolated bound state, with weak hybridization between the impurity sites. Given this weak hybridization, the bands arising from the interplay between different sites have a small bandwidth, such that all bound state energies $E$ remain of the order of $E_\pm$. Using this reasoning, we can now rewrite the Schr\"odinger equation for small energies $|E|\sim|E_\pm|\ll |M|$ to find an effective low-energy tight-binding model for the bound states associated with the dilute impurities. Rearranging the Schr\"odinger equation evaluated at impurity site $l$ such that all on-site terms appear on the left-hand side and all hybridization terms are on the right-hand side, we have
\begin{equation}
\big[\mathbb{1}-V_M\hat{\mathcal{G}}_0^+(0;E)\sigma^x\big]\psi(x_l)=V_M\sum_{m\neq l}\hat{\mathcal{G}}_0^+(x_{lm};E)\sigma^x\psi(x_m),
\end{equation}
where $x_{lm}=x_l-x_m$. This expression is still exact, but difficult to solve exactly in the current form. The next step is therefore to consider the case $|E_\pm|\ll |M|$ in the dilute limit and expand the above equation to lowest nontrivial order in $E$ around zero. For the left-hand side, this is the linear order. On the right-hand side, however, we can simply set $E\rightarrow 0$: the hybridization terms are already small by virtue of the dilute limit, such that the linear order term in $E$ can be interpreted as higher order. Performing this Taylor expansion and rearranging the result, we find
\begin{equation}
E\psi(x_l)\approx\sum_m\hat{H}_{lm}\psi(x_m),
\end{equation}
with
\begin{align}
\hat{H}_{lm}=-M\bigg[\bigg(&1+\frac{2\hbar v_F}{V_M}\text{sgn}(M)\delta_{l,m}\bigg)\sigma^x\nonumber\\
&+i\,\text{sgn}(M)\text{sgn}(x_{lm})\sigma^z\bigg]e^{-|x_{lm}|/\xi_0}.\label{eq:Hlm}
\end{align}
We have thus shown that the low-energy bound state bands can indeed be described by an effective tight-binding model, as long as the system is in the dilute limit, and the impurity strength has been tuned such that $|V_M|\sim 2\hbar v_F$. Moreover, in this limit the on-site energy from the tight-binding model,
\begin{equation}
E_0=-M\left[1+\frac{2\hbar v_F}{V_M}\text{sgn}(M)\right],
\end{equation}
coincides with $E_+$ (provided that the sign of $V_M$ is chosen opposite to the sign of $M$), while all hopping terms are exponentially suppressed. This is consistent with our previous assessment of the situation, confirming the validity of the approximations.

\subsection{Phase diagram}
\begin{figure}[t]
\centering
\includegraphics[width=0.45\textwidth]{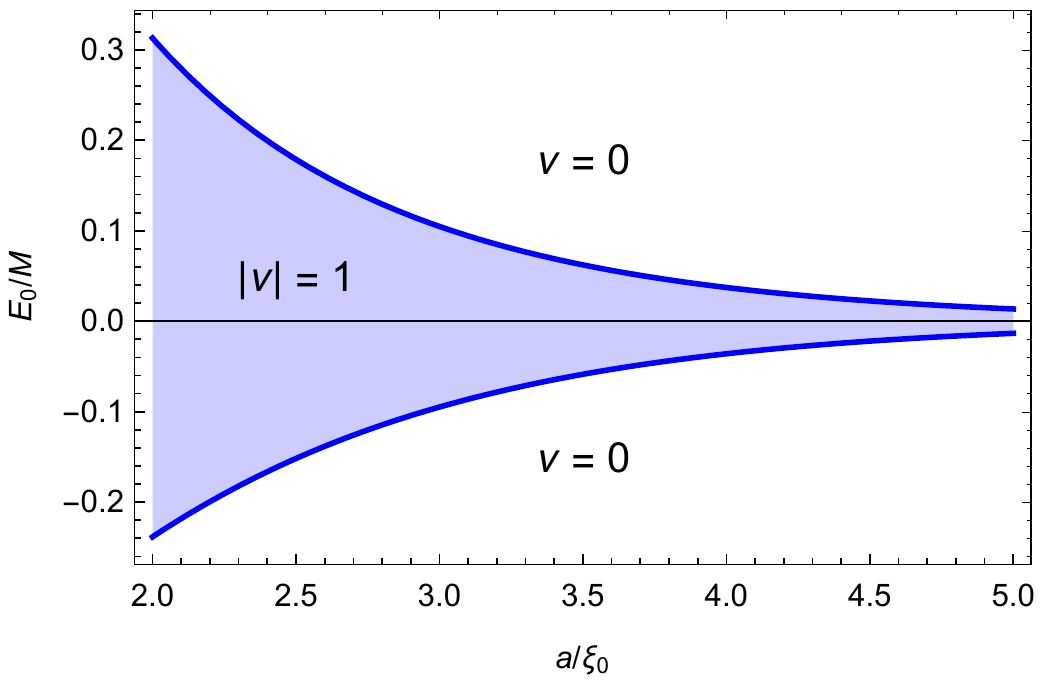}
\caption{\label{fig:phasediagram}Phase diagram of the effective low-energy tight-binding model of impurity bound states in the dilute limit. The phase diagram reveals the existence of a large topological phase with nontrivial winding number $\nu$. Note that the region $a/\xi_0<2$ has been omitted because of the fact that the exploited approximations break down when $a$ becomes of the order of $\xi_0$.}
\end{figure}

In order to examine the properties of the bands emerging from the effective tight-binding model from Eq.~(\ref{eq:Hlm}), we now consider a homogeneous distribution of impurities, \textit{i.e.}, the impurities are arranged on a chain with lattice constant $a=a_\text{imp}$. Also imposing periodic boundary conditions, this allows us to Fourier transform the real-space matrix $\hat{H}_{lm}$ to momentum space, resulting in the following Bloch Hamiltonian:
\begin{align}
\hat{H}(k)&=\left[E_0+M\left(1+\frac{\sinh(a/\xi_0)}{\cos(ak)-\cosh(a/\xi_0)}\right)\right]\sigma^x\nonumber\\
&\quad\,-|M|\frac{\sin(ak)}{\cos(ak)-\cosh(a/\xi_0)}\sigma^z,\label{eq:Hk}
\end{align}
which has the form discussed above for a one-dimensional system in class AIII of the Altland-Zirnbauer classification~\cite{AltlandZirnbauer}. The corresponding ${\mathbb Z}$ topological invariant is the winding number, or chiral charge:
\begin{equation}
\nu=\frac{1}{2\pi}\int\limits_{-\pi/a}^{\pi/a}\mathrm{d}k\,\frac{h_x(k)\partial_kh_z(k)-h_z(k)\partial_kh_x(k)}{|\mathbf{h}(k)|^2},
\end{equation}
counting the number of times the vector $\mathbf{h}(k)$ (with components $h_{x,z}(k)$) loops around the origin when $k$ goes from $-\pi/a$ to $\pi/a$. For our tight-binding model, we find that $\nu=\text{sgn}(M)$ if
\begin{equation}
\frac{\sinh(a/\xi_0)}{\cosh(a/\xi_0)+1}-1<E_0/M<\frac{\sinh(a/\xi_0)}{\cosh(a/\xi_0)-1}-1,
\end{equation}
and $\nu=0$ otherwise, corresponding to a topological phase and a trivial phase, respectively. The resulting phase diagram is shown in Fig.~\ref{fig:phasediagram}. Note that the topological phase extends all the way to $a\rightarrow\infty$ along the line $E_0=0$. We furthermore emphasize that we observe only two values for $|\nu|$. The fact that higher winding numbers do not emerge for our particular setup is a consequence of its simplicity, such as the linear edge dispersion from Eq.~(\ref{eq:Hmin}). However, the topological invariant is nevertheless of type ${\mathbb Z}$, and can in principle take any integer value for similar impurity setups with the same symmetries.

\subsection{Boundary modes of the impurity bound states}
\begin{figure}[t]
\centering
\includegraphics[width=0.45\textwidth]{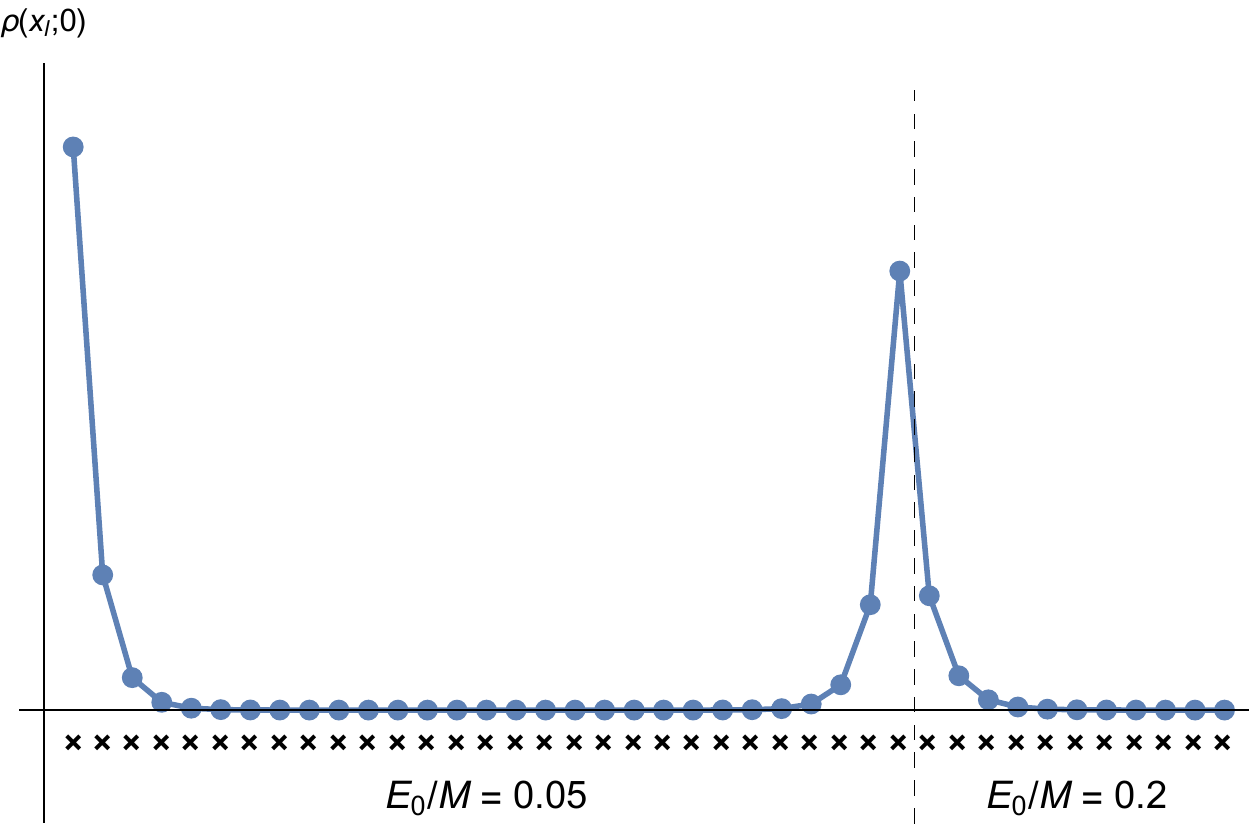}
\caption{\label{fig:doseff}Zero-energy local density of states $\rho(x_l;0)$ at the impurity sites of a finite chain with $a/\xi_0=3$. The chain consists of a part with $E_0/M=0.05$ (to the left of the dashed line, corresponding to $|\nu|=1$), and a part with $E_0/M=0.2$ (to the right of the dashed line, with $\nu=0$). The crosses at the bottom denote the positions of the impurities. This density of states profile confirms the presence of localized zero-energy modes, located both at the end of a chain in the topological phase, and at the boundary between two regions that are respectively in the topological and in the trivial phase.}
\end{figure}

\begin{figure}[t]
\centering
\includegraphics[width=0.45\textwidth]{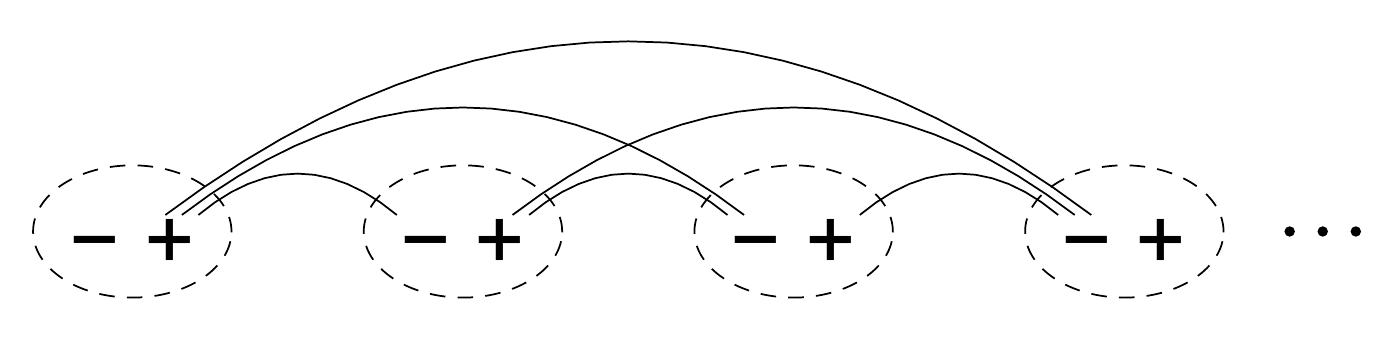}
\caption{\label{fig:hopping}Schematic representation of the hopping elements of the effective low-energy tight-binding Hamiltonian for $M>0$. Each ``$-\;+$'' pair denotes the two $\sigma^y$-eigenstates at an impurity site, while the lines indicate the states that are connected to each other. Note that the left and right ends of the chain each host a state that is not connected to any other impurity site, such that these modes are only directly involved in the on-site hybridization.}
\end{figure}

The nontrivial topology of the region with $\nu\neq 0$ indicates the presence of topologically protected zero-energy boundary modes. These modes are expected to emerge both at the end points of a finite impurity chain in the topological phase, and at boundaries between regions with different winding numbers. Fig.~\ref{fig:doseff} shows a plot of the local density of states (LDOS) $\rho(x;E)\equiv\sum_n\delta(E-E_n)|\psi_n(x)|^2$ for a finite chain that contains a domain wall, demonstrating that the expected boundary modes are indeed present. In order to understand the origin of the boundary modes and analytically prove their presence for certain parameter values, we return to the effective tight-binding model from Eq.~(\ref{eq:Hlm}) and perform a unitary transformation. In particular, we change the basis of the Hamiltonian from the eigenstates of $\sigma^z$ to the eigenstates of $\sigma^y$, \textit{i.e.}, the component of the spin that does not appear in the original Hamiltonian due to the approximate chiral symmetry. In the language of second quantization, this gives
\begin{equation}
H=\sum_{lm}c_l^\dagger\hat{H}^\prime_{lm}c^{\phantom{\dagger}}_m,\qquad c_l=\begin{pmatrix}c_{l+}\\ c_{l-}\end{pmatrix},
\end{equation}
where
\begin{equation}
\hat{H}^\prime_{lm}=\hat{U}\hat{H}_{lm}\hat{U}^\dagger,\qquad \hat{U}=\frac{1}{\sqrt{2}}\begin{pmatrix}1 & -i\\ 1 & i\end{pmatrix},
\end{equation}
and $c_{l\pm}^\dagger$ ($c_{l\pm}$) are the creation (annihilation) operators corresponding to the eigenstates of $\sigma^y$ at impurity site $l$. Considering $M>0$ and labeling the impurities such that $x_m>x_l$ if $m>l$, the above can explicitly be written as
\begin{align}
H&=E_0\sum_lc_l^\dagger\sigma^yc^{\phantom{\dagger}}_l\nonumber\\
&\,+\left[2i|M|\sum_l\sum_{m>l}c_l^\dagger\begin{pmatrix}0 & e^{-|x_{lm}|/\xi_0}\\ 0 & 0\end{pmatrix}c^{\phantom{\dagger}}_m+h.c.\right].\label{eq:Hdimer}
\end{align}
The case $M<0$ leads to the same result, but with the matrix from the second line being transposed. We thus observe that many of the hopping terms of the tight-binding Hamiltonian vanish: if $M>0$ ($M<0$), then a ``$+$'' mode at a given impurity site is only coupled to all ``$-$'' modes living on impurity sites to its right (left), and conversely a ``$-$'' mode is only coupled to all ``$+$'' modes to its left (right), see Fig.~\ref{fig:hopping}. Consequently, each end of a finite chain hosts a degree of freedom that is not connected to any other impurity, therefore only appearing in the on-site term that is governed by $E_0$. In the special case $E_0=0$ (corresponding to the horizontal axis of Fig.~\ref{fig:phasediagram}), these particular $\sigma^y$ eigenstates completely decouple from the system, such that they are indeed the anticipated zero-energy boundary modes. For more general parameter values, the presence or absence of these boundary modes is determined by whether or not the on-site hybridization is strong enough to delocalize them by ``injecting'' them into the bands.

The above analysis leads to several important implications. First, we observe that the decoupling of boundary modes that happens at $E_0=0$ is completely independent of the actual configuration of the impurities. Indeed, the configuration does not influence which of the hopping elements are zero, and instead only determines the strength of the nonzero components, see Eq.~(\ref{eq:Hdimer}). The boundary modes therefore emerge even in the absence of any spatial symmetry along the chain, such as the mirror line symmetry $\mathcal{M}_x$ (we assume that the host does not react to the impurities, meaning that $\mathcal{M}_x$ is preserved for the host). Moreover, it should be noted that the boundary modes survive disorder in the form of a set of random local potentials, provided that the boundary modes are not shifted into the impurity bands (acting as the ``bulk bands'' of this system). This further demonstrates the robustness of the boundary modes, while confirming that the setup does also not rely on overall particle-hole symmetry (although it is required for the host). Instead, the emergence of the boundary modes is a consequence of the antiunitary $\mathcal{M}_z\Theta$ symmetry: together with $\mathcal{M}_x$, this symmetry is responsible for the approximate chiral symmetry of the effective tight-binding model, resulting in the absence of $\sigma^y$ in $\hat{H}_{lm}$. This in turn allows for the large amount of vanishing hopping elements in Eq.~(\ref{eq:Hdimer}). However, when the configuration of the impurities violates the mirror line symmetry $\mathcal{M}_x$, the absence of $\sigma^y$ is no longer guaranteed. As such, any boundary modes that remain upon breaking $\mathcal{M}_x$ are not protected by symmetry and are therefore specific to the model. Finally, we note that our effective model bears close resemblance to the Su-Schrieffer-Heeger (SSH) chain~\cite{SSH} in the even sector, where the role of the sublattices is now played by the two spin states. This is consistent with this section's observation that the boundary modes are spin polarized.

\subsection{Quantized excess charges protected by mirror symmetry}
Having explicitly demonstrated the emergence of boundary modes assuming the quantum spin Hall insulator preserves the planar mirror symmetry ${\mathcal M}_z$ in the absence of the external planar magnetic field, we next consider the situation in which a, {\it e.g.}, Rashba spin-orbit coupling breaks ${\mathcal M}_z$. Consequently, the combined antiunitary ${\mathcal M}_z \Theta$ symmetry is broken as well, such that we are left with an effective one-dimensional system where only a mirror line is preserved. This also implies that the effective one-dimensional Hamiltonian possesses all three Pauli matrices and reads
\begin{figure}[t]
\centering
\includegraphics[width=0.5\textwidth]{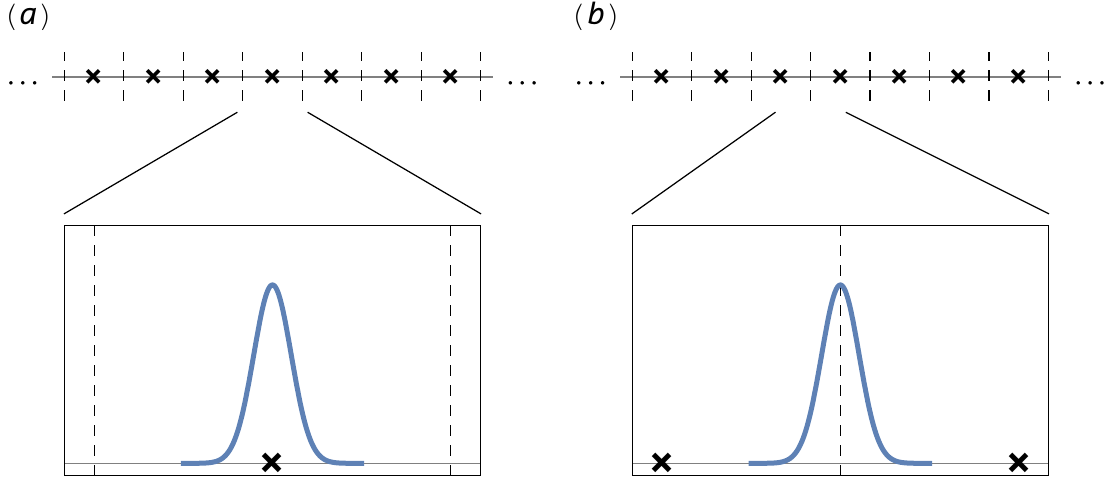}
\caption{\label{fig:local}Possible locations for the bulk electrons in a given unit cell for a half-filled setup with mirror symmetry. The crosses denote the impurity sites, the dashed lines represent the borders between different unit cells, and the Gaussian curves represent the Wannier wave functions. Note that these Wannier functions must be the same for each unit cell due to translational symmetry, and that they must be either even or odd with respect to their central point. $(a)$: Wannier function located at an impurity site, associated with mirror eigenvalues $m_0=m_\pi$ (trivial phase). $(b)$: Wannier function located between two impurity sites, corresponding to mirror eigenvalues $m_0=-m_\pi$ (topological phase).}
\end{figure}
\begin{equation}
\hat{H}(k)=h_0(k)\mathbb{1}+h_x(k)\sigma^x+h_y(k)\sigma^y+h_z(k)\sigma^z,\label{eq:Hgen}
\end{equation}
where, as before, we can neglect the identity term, whereas the functions $h_{y,z}(k)$ are odd in momentum $k$. Since this 1D model is in symmetry class A (inherited from the microscopic model from which it is derived), it generally does not have a topological invariant protected by an approximate internal symmetry. However, the presence of the mirror line ${\mathcal M}_x$ allows one to define a crystalline topological invariant regulating the appearance of fractional excess charges $Q_{b} = 0, 1/2 \mod 1$ appearing at the end of the one-dimensional impurity chain. To show how this fractional corner charge can be read off from the bulk properties of the system, we recall that given that the bulk is an insulator, the many-body ground state of the system can be represented in terms of exponentially localized Wannier functions. Note that this is strictly valid only in the one-dimensional settings we are concerned with in this work. Looking at a single unit cell, the mirror symmetry furthermore dictates that for an odd number of electrons in the unit cell the corresponding Wannier function must either be centered at an impurity site $(1a)$ or centered in the middle between two neighboring impurity sites $(1b)$, see Fig.~\ref{fig:local}. These are the two maximal Wyckoff positions in the unit cell, whose site symmetry group contains the mirror symmetry~\cite{mie18}. The fact that the stabilizer group of $1a$ is only generated by the mirror symmetry, {\it i.e.}, $\left\{{\mathcal M}_x | 0\right\}$, whereas the stabilizer group of $1b$ also contains a translation of the unit cell $\left\{ {\mathcal M}_x | 1 \right\}$, leads to the induced elementary band representation for exponentially localized Wannier functions listed in Table~\ref{tab:tab1d}. It now follows that the location of the electrons in each unit cell is directly related to the eigenvalues of the mirror operator $\mathcal{M}_x=i\sigma^x$ at the special points $k=0$ and $k=\pi/a$: the electrons are localized at the impurity sites if the mirror eigenvalues of the filled band at the special points satisfy $m_0=m_\pi$, and they are located in between impurity sites if $m_0=-m_\pi$. In the latter case, opening the chain results in the emergence of a half-integer fractional corner charge whose quantized value is protected by the mirror symmetry.
\begin{table}[t]
\caption{Elementary band representation for the one-dimensional space group of a mirror-symmetric chain. The first column indicates the maximal Wyckoff positions, the second column the corresponding induced band representation, and the last two columns the mirror eigenvalues at the center and edge of the 1D Brillouin zone.}
\label{tab:tab1d}
\begin{tabular}{|c|c|c|c|}
\hline
Wyckoff position & Representation & $\Gamma$ & $X$ \tabularnewline
\hline
$1a$ & $\rho_{i}^{1a} \uparrow {\mathcal G}$ & $i$ & $i$  \tabularnewline
& $\rho_{-i}^{1a} \uparrow {\mathcal G}$ & $-i$ & $-i$  \tabularnewline
\hline
$1b$ & $\rho_{i}^{1b} \uparrow {\mathcal G}$ & $i$ & $-i$  \tabularnewline
& $\rho_{-i}^{1b} \uparrow {\mathcal G}$ & $-i$ & $i$  \tabularnewline
\hline
\end{tabular}
\end{table}

Applying the above to the mirror-symmetric setup from Eq.~(\ref{eq:Hk}), we see that the trivial phase $\nu=0$ indeed corresponds to $m_0=m_\pi$, while the topological phase $|\nu|=1$ corresponds to $m_0=-m_\pi$. The origin of the boundary modes discussed in the preceding section can be therefore seen as a consequence of an excess charge - boundary mode correspondence that is realized in the additional presence of the chiral symmetry.

\section{Numerical analysis}\label{sec:numerical}
Motivated by the results from the dilute limit, we will now numerically investigate the possibility of similar behavior away from this limit, using the full microscopic lattice model from Eqs.~(\ref{eq:Hfull})--(\ref{eq:Himplatt}). The rapidly increasing computation times when working with large matrices require working with small system sizes. However, in order to reduce finite size effects, it is necessary that the dimensions of the system are still larger than the correlation length $\xi$. Recalling that the correlation length typically goes like $\xi\sim 2\hbar v_F/\Delta_\text{edge}$, this can be done by choosing the magnetic field $B$ sufficiently large. However, since we are specifically interested in the interactions between the gapped edge and the impurity bound states, it is also necessary to keep the edge gap much smaller than the bulk gap, \textit{i.e.}, $|B|\ll |t|$. The difficulty in numerically solving the microscopic model is therefore to tune $B$ such that it is small compared to $t$, while choosing it large enough such that the minimum system size is still feasible. One parameter set that achieves this is given by $t_2/t=B/t=0.1$, which we will use throughout this section. It should, however, be noted that the requirement to fine-tune the parameters is only a practical limitation, not a physical one.
\begin{figure}[t]
\centering
\includegraphics[width=0.5\textwidth]{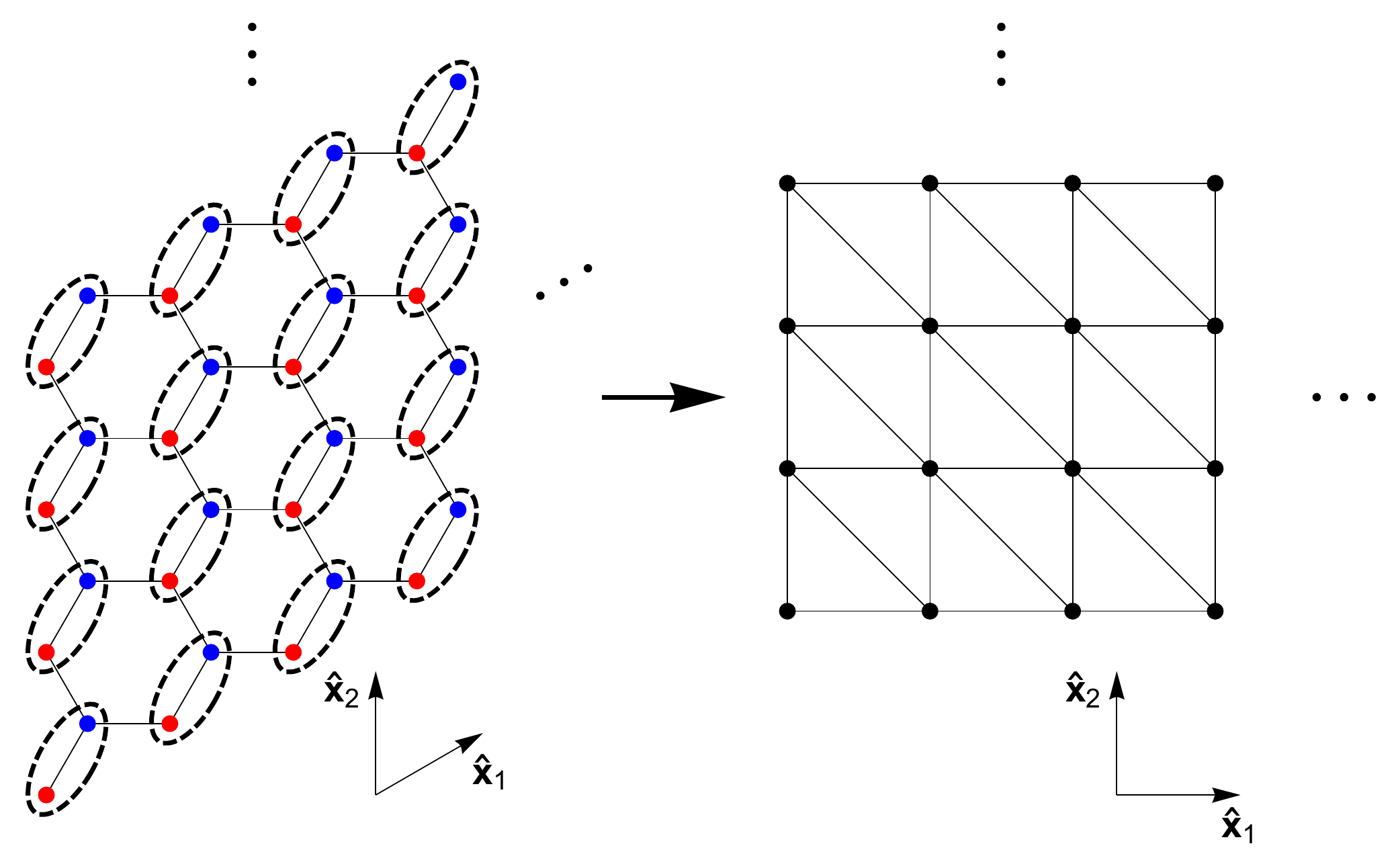}
\caption{\label{fig:setup}Illustration of the numerical implementation of the microscopic lattice model, consisting of a mapping of the honeycomb lattice to a square lattice. The left panel shows the original (finite) honeycomb lattice, with double zigzag edges and lattice vectors $\hat{\mathbf{x}}_{1,2}$. Each unit cell consists of two lattice sites, while each lattice site in turn supports two spin states. Every unit cell is therefore represented by a $4\times 4$ matrix. The lattice is then deformed in such a way that the lattice vectors become orthogonal, leading to a square lattice of these unit cells (right panel). The bonds between unit cells (denoted by the lines in the right panel) are represented by $4\times 4$ matrices as well, the elements of which contain the original nearest neighbor and next-nearest neighbor bonds.}
\end{figure}

\begin{figure}[t]
\centering
\includegraphics[width=0.45\textwidth]{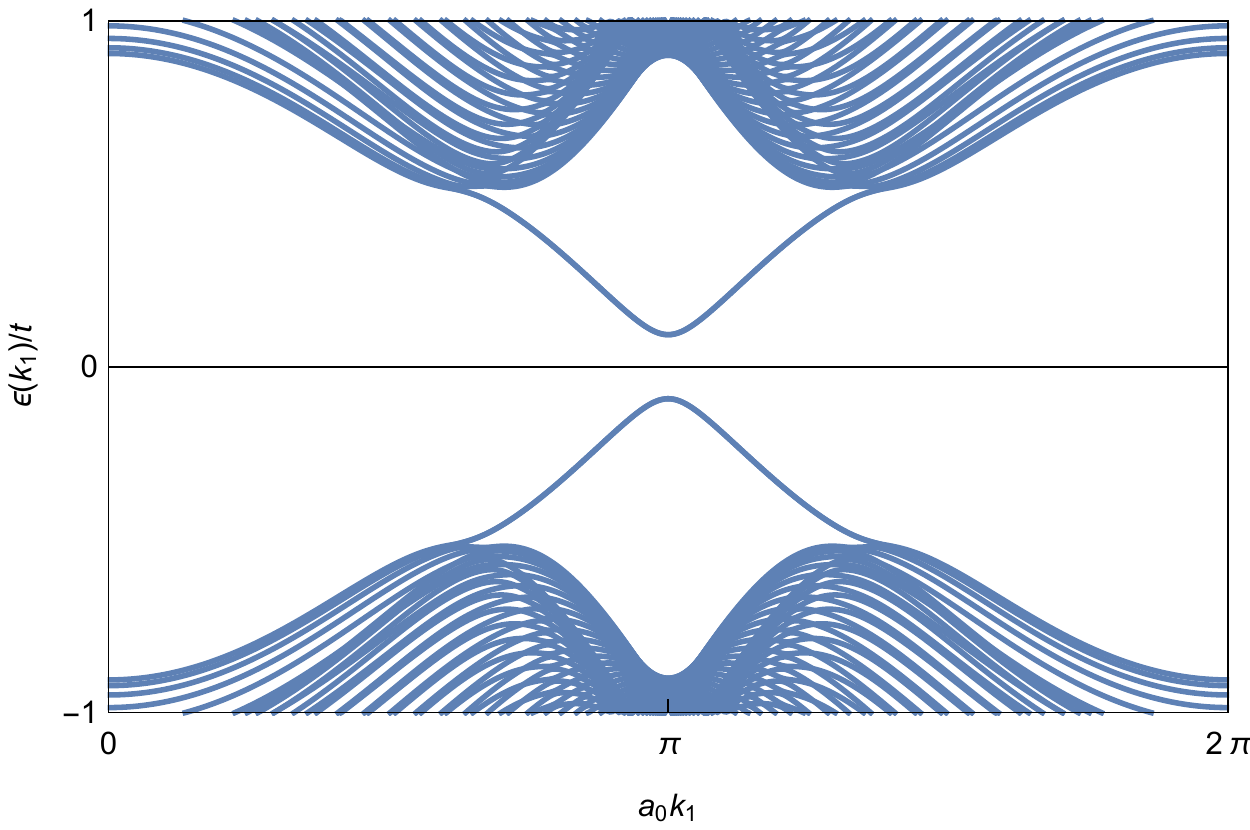}
\caption{\label{fig:bandstruc}Band structure of the microscopic model in absence of impurities, with parameter values $t_2/t=B/t=0.1$, $N_2=40$ unit cells in the $\hat{\mathbf{x}}_2$ direction, and periodic boundary conditions in the $\hat{\mathbf{x}}_1$ direction. The gapped cone is strongly localized at the zigzag edge of the cylindrical geometry. The associated edge gap is indeed much smaller than the bulk gap, while the low-energy dispersion relation of the gapped cone closely resembles that of the effective model from Eq.~(\ref{eq:Hmin}).}
\end{figure}

We consider a finite honeycomb lattice with zigzag edges in both directions. As demonstrated in Fig.~\ref{fig:setup}, this structure is implemented numerically by mapping the honeycomb lattice structure to that of a square lattice. Since each unit cell consists of two orbitals and two spin states, this translates to a $4N\times 4N$ Hamiltonian matrix, where $N$ is the number of unit cells. The $4\times 4$ blocks on the diagonal of this matrix consist of all on-site energies (including the local perturbations that represent the impurities) and the hoppings within a single unit cell, while all other $4\times 4$ blocks contain the hoppings between different unit cells. Using this implementation, we first extract the Fermi velocity and the size of the edge gap in order to later compare the numerical results to the results from Sec.~\ref{sec:dilute}. To do so, we take the clean system and impose periodic boundary conditions in one of the directions, allowing for a Fourier transformation of that direction to momentum space. The resulting matrix is then diagonalized to find the dispersion relation $\epsilon(k)$, shown in Fig.~\ref{fig:bandstruc}. For $t_2/t=B/t=0.1$, fitting the edge modes of this band structure to the low-energy boundary model from Eq.~(\ref{eq:Hmin}) gives the values $\hbar v_F/a_0t=0.55$ and $M/t=0.093$. For later reference, we note that plugging these numbers into the effective model from Eq.~(\ref{eq:Hmin}) implies that the low-energy correlation length is of the order $\xi_0\sim 6a_0$. Moreover, an isolated impurity bound state is expected to have an energy close to the center of the gap (\textit{i.e.}, $|E_0/M|\ll 1$) if the impurity strength is chosen according to $\tilde{V}_M/t=-1.1$.

\begin{figure*}[t]
\centering
\includegraphics[width=\textwidth]{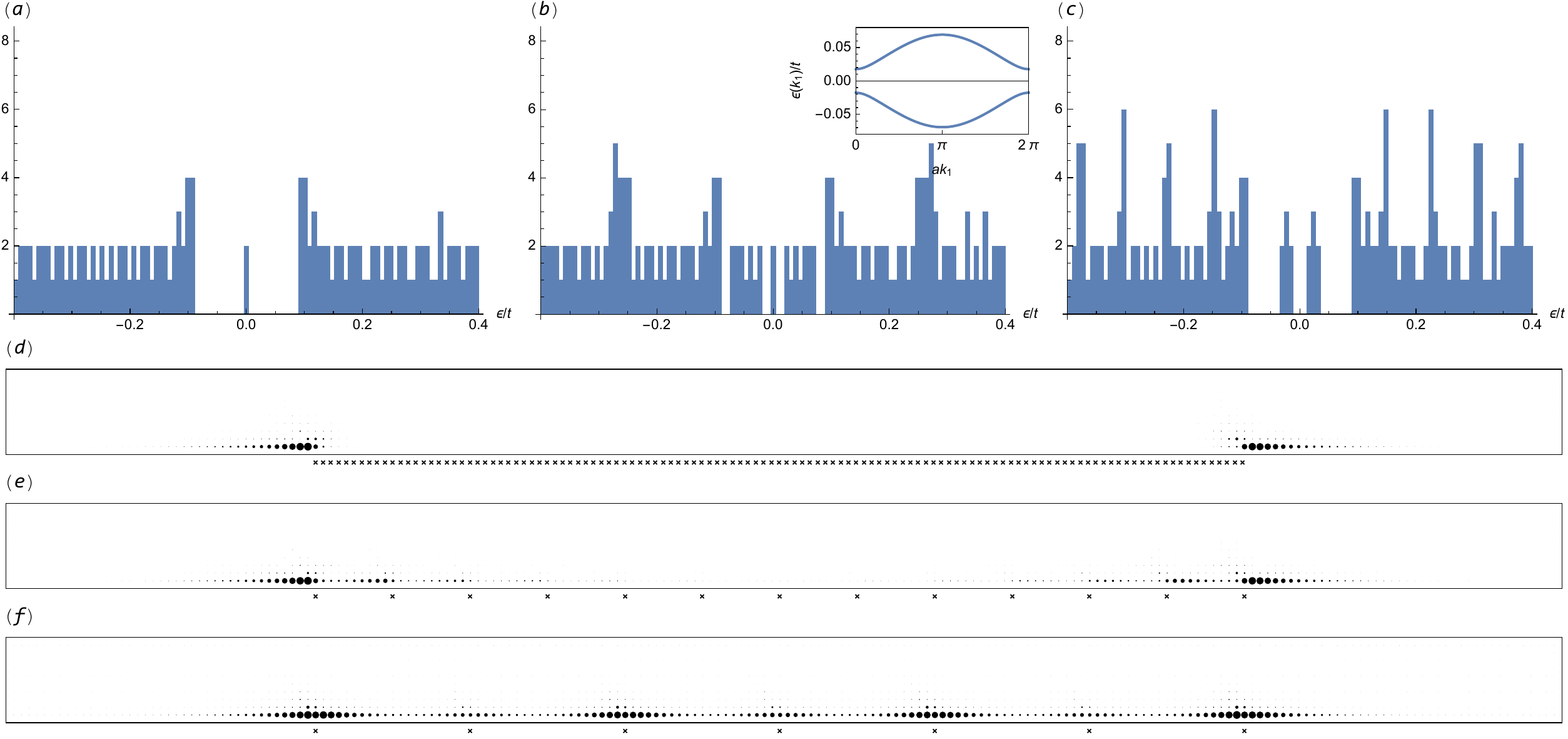}
\caption{\label{fig:results}$(a)-(c)$: Histograms of the eigenvalues of the microscopic lattice Hamiltonian~(\ref{eq:Hfull}) in the presence of a chain of magnetic impurities along the bottom edge, for different impurity lattice constants $a$. The dimensions of the lattice are $N_1=200$ by $N_2=10$ unit cells, and the parameter values are $t_2/t=B/t=0.1$ and $\tilde{V}_M/t=-1.1$. Panel $(a)$ shows the dense limit $a/a_0=1$ (\textit{i.e.}, each unit cell in the chain hosts an impurity), while panels $(b)$ and $(c)$ correspond to the more dilute configurations $a/a_0=10$ and $a/a_0=20$, respectively. The inset of panel $(b)$ furthermore shows the dispersion of the bound state bands in the case of periodic boundary conditions in the $\hat{\mathbf{x}}_1$ direction. Note that the histograms are proportional to the total density of states $\int\mathrm{d}x\,\rho(x;E)$. $(d),(e)$: Zero-energy LDOS profiles corresponding to the zero-energy peaks from panels $(a),(b)$. Each dot represents a unit cell, with the size of each dot being proportional to the corresponding LDOS, scaled to the largest value. The small crosses below each profile indicate which unit cells are hosting the impurities. Comparing panel $(e)$ to panel $(d)$, we see that the observed zero-energy boundary modes become less well localized as the impurity density is decreased. $(f)$: LDOS profile for the energy of an isolated bound state, $E/M=-0.23$, in the dilute case $a/a_0=20$. This profile clearly reveals the presence of weakly hybridized bound states on each of the impurity sites.}
\end{figure*}

With the above in mind, we now take open boundaries in all directions, while applying a strong local potential to the corner sites to eliminate the effects of any eventual corner states. The impurities are included by applying Eq.~(\ref{eq:Himplatt}) to the outermost lattice site of selected unit cells along the bottom edge of the system (we have also checked that with other sublattice structures the results are not qualitatively changed). In the case of a single impurity with $\tilde{V}_M/t=-1.1$, this results in bound state energies of $E_\pm/M=\pm 0.23$, which can in turn be used to identify $|E_0/M|\approx 0.23$. The case of an impurity chain is furthermore studied by taking an elongated system of $N_1=200$ by $N_2=10$ unit cells, allowing us to make the chain sufficiently long for the corresponding finite size effects to become small. The resulting energy eigenvalues and LDOS profiles are shown in Fig.~\ref{fig:results} for several different impurity lattice constants $a$ (note that the small particle-hole asymmetry in Fig.~\ref{fig:results} stems from the strong local corner potential).

The limit opposite to the setup discussed in Sec.~\ref{sec:dilute} is the limit with the largest possible density of impurities in the chain. In the language of the microscopic model, this dense limit is given by $a/a_0=1$, which translates to $a/\xi_0\approx 0.17$ for our choice of parameters. As shown in panels $(a)$ and $(d)$ of Fig.~\ref{fig:results}, the dense limit still supports the boundary modes at the ends of the chain that were also found in the dilute limit. Additionally, we find that these modes can still be moved away from zero energy by means of a local potential on their respective sites. This confirms that the behavior that was revealed by the tight-binding model from Eq.~(\ref{eq:Hlm}) does in fact extend beyond the dilute limit, despite the breakdown of the approximations used in its derivation. As a particular example of this breakdown, the histogram of energy eigenvalues reveals that the bound state bands no longer reside in the edge gap, meaning that they have moved into the edge bands. The fundamental assumption that all bound state energies remain close to the center of the gap is therefore no longer valid. This is a direct consequence of the strong hybridization between the individual impurity bound states. On the other hand, we also consider the more dilute arrangements $a/a_0=10$ and $a/a_0=20$, corresponding to $a/\xi_0\approx 1.7$ and $a/\xi_0\approx 3.3$, respectively. Combining these values with the observation from panels $(b)$ and $(c)$ of Fig.~\ref{fig:results} that the bound state bands are now entirely contained within the edge gap, we expect the tight-binding model to be a good approximation. Furthermore comparing these values for $a/\xi_0$ and $|E_0/M|\approx 0.23$ to the phase diagram from Fig.~\ref{fig:phasediagram}, we predict $a/a_0=10$ to be in the topological phase, and $a/a_0=20$ in the trivial phase. This is indeed confirmed by panels $(b)$, $(c)$ and $(e)$ of Fig.~\ref{fig:results}: the case $a/a_0=10$ has zero-energy modes that are localized at the ends of the chain, while $a/a_0=20$ does not have any zero modes. Moreover, the bound state energies of the latter reside in a narrow range around the isolated bound state energies $E_\pm$, such that this case can be identified as dilute. Panel $(f)$ finally shows the LDOS profile corresponding to $a/a_0=20$ for $E=E_-$; the profile implies that the bound states can indeed be approximated as a series of isolated bound state solutions. Altogether, the results from Fig.~\ref{fig:results} thus successfully bring together the approximate effective low-energy tight-binding model and the underlying microscopic lattice model, confirming the validity of the former and extending its qualitative behavior all the way to the dense limit.

\section{Conclusion and discussion}\label{sec:conclusion}

Starting from a two-dimensional quantum spin Hall insulator, we constructed a one-dimensional insulator of the AIII class. After applying a magnetic field parallel to the boundary, the gapped boundary theory has an $\mathcal{M}_x$ mirror symmetry as well as an antiunitary symmetry $\mathcal{M}_z\Theta$. This combination guarantees the chiral symmetry of the (bulk) boundary theory. We then deposited magnetic impurities onto the boundary. The bound states of these impurities hybridize and disperse along the edge. This effective tight-binding model is part of the AIII symmetry class and itself has zero-dimensional boundary modes together with half-integer boundary excess charges. It is interesting to compare our lattice model, Eq.~\eqref{eq:Hdimer}, to the more well-known SSH model. While in the SSH model the boundary charges depend on the cutting scheme and show an even-odd effect, this effect is absent in the model presented here. Irrespective of the boundary there is a boundary charge in the nontrivial phase. We determined a full phase diagram based on the effective low-energy model, Eq.~\eqref{eq:Hdimer}, which was successfully checked numerically against the full microscopic lattice model. An attractive feature of the construction presented here is that while the $\mathcal{M}_x$ symmetry is important for the bulk theory, it is not for the boundary. This implies that the results presented are robust to disorder and spatial variations of the impurity sites. We also showed that in the presence of Rashba coupling, which changes the symmetry class to A, one can still expect localized boundary excess charges although the topological index is zero. These modes are protected by the mirror symmetry of the bulk theory.

For the future it is an interesting question to what extent one can engineer further states using this pollution technique where you use a higher dimensional nontrivial host system as a starting point.

\section*{Acknowledgements}
This work is part of the D-ITP consortium, a program of the Netherlands Organisation for Scientific Research (NWO) that is funded by the Dutch Ministry of Education, Culture and Science (OCW). C.O. acknowledges support from a VIDI grant (Project 680-47-543) financed by NWO.


\end{document}